\documentclass{article}

\usepackage{arxiv}

\usepackage[utf8]{inputenc} 
\usepackage[T1]{fontenc}    
\usepackage{hyperref}       
\usepackage{url}            
\usepackage{booktabs}       
\usepackage{amsfonts}       
\usepackage{nicefrac}       
\usepackage{microtype}      
\usepackage{graphicx}

\title{Recurrent Neural Network-based Model for Accelerated Trajectory Analysis in AIMD Simulations}

\author{
  Mohammad Javad~Eslamibidgoli$^{*}$ \\
  School of Computing Science\\
  Simon Fraser University\\
  8888, University Drive \\
    Burnaby, BC, Canada, V5A 1S6 \\
  \texttt{meslamib@sfu.ca} \\
   \And
 Mehrdad Mokhtari \\
  Department of Chemistry\\
  Simon Fraser University\\
  8888, University Drive \\
  Burnaby, BC, Canada, V5A 1S6 \\
    \texttt{mokhtari@sfu.ca} \\
  \And
 Michael H.~Eikerling \\
IEK-13, Institute of Energy \\
and Climate Research,\\
Forschungszentrum Jülich GmbH, \\
52425 Jülich, Germany \\
    \texttt{m.eikerling@fz-juelich.de} \\
}

\begin{document}
\maketitle

\begin{abstract}
The presented work demonstrates the training of recurrent neural networks (RNNs) from distributions of atom coordinates in solid state structures that were obtained using \textit{ab initio} molecular dynamics (AIMD) simulations. AIMD simulations on solid state structures are treated as a multi-variate time-series problem. By referring interactions between atoms over the simulation time to temporary correlations among them, RNNs find patterns in the multi-variate time-dependent data, which enable forecasting trajectory paths and potential energy profiles. Two types of RNNs, namely gated recurrent unit and long short-term memory networks, are considered. The model is described and compared against a baseline AIMD simulation on an $ Ir_{0.75}W_{0.25}O_{2} (100) $ slab. Findings demonstrate that both networks can potentially be harnessed for accelerated statistical sampling in computational materials research.
\end{abstract}


\section{Introduction}

Although the primary methods were developed in the 1950s to 80s, breakthroughs in machine learning have occurred only recently, with many interesting applications such as in computer vision ~\cite{voulodimos2018deep}, speech and language technology ~\cite{gaikwad2010review}, self-driving cars~\cite{bojarski2016end}, recommender systems~\cite{DBLP:journals/corr/ChengKHSCAACCIA16}, financial predictions~\cite{wilson1994bankruptcy}, and many more~\cite{cho2014learning}. This is mainly owed to recent advances in neural network architectures and algorithms~\cite{schmidhuber2015deep}, escalating growth of data for model training~\cite{zikopoulos2011understanding}, powerful parallel computer processing, enhanced frameworks for implementation~\cite{paszke2017automatic} and, of course, larger involvement from the scientific community as well as industry in the field.   

Machine learning is well-established in areas like bioinformatics as a computational method to analyze and interpret biological data~\cite{chicco2017ten, wei2018prediction}. Deep neural network architectures such as multi-layered networks, convolutional neural networks (CNNs), recurrent neural networks (RNNs) and memory networks are the most used architectures in this area~\cite{hou2017deepsf}. Machine learning has also shown to be promising in theoretical chemistry~\cite{aspuru2018matter} mainly to speed up the discovery of novel compounds or materials~\cite{sanchez2018inverse, gomez2018automatic, huan2017universal, li2017machine}, as well in terms of predicting the potential energy surface of chemical structures~\cite{brockherde2017bypassing, chmiela2017machine}.  

Extensive research is focused on developing inter-atomic potentials~\cite{kassal2008polynomial} or performing AIMD simulations~\cite{deringer2018realistic, gastegger2017machine}. Brockherde \textit{et al.} employed the kernel ridge regression training algorithm based on an external Gaussian potential to generate the machine learned potential-density map of a given molecule~\cite{brockherde2017bypassing}; their proposed approach starts with the construction of descriptors, which encode the structure of each molecule from a training data-set. The kernel matrix is then developed from a kernel function to represent the correlation of descriptors to each other~\cite{brockherde2017bypassing}. The data-set in their work is composed of 2000 DFT optimized structures. 

Gomez-Bombarelli \textit{et al.}~\cite{gomez2018automatic} have recently used variational autoencoders to transform the discrete representation of molecular structures into a continuous latent space, from which a decoder neural network converts these continuous representations back into a discrete molecular representation. This approach allows designing new molecules and enables efficient exploration and optimization of the chemical spaces of compounds~\cite{gomez2018automatic}. Chmiela \textit{et al.} advanced a symmetrized gradient-domain machine learning model, in which one can create a molecular force field of intermediate-sized organic compounds with the same accuracy as high-level \textit{ab initio} calculations only using limited samples (of size 1000) of AIMD calculated trajectories~\cite{chmiela2019sgdml}. In another work, Xie and Grossman proposed a crystal graph convolutional neural network framework to provide a universal representation of crystal materials and an accurate prediction of DFT calculated properties of various crystal structures and compositions~\cite{xie2018crystal}.

This paper pioneers the use of RNN-based models in predicting trajectory paths and potential energy profiles of solid-state structures in AIMD simulations. The main purpose of AIMD simulation is to generate as many as possible statistically independent configurations of the system under study at affordable computational costs. This requires extensive simulation runs and sufficiently long intervals between configurations which are used in thermodynamic analyses to ensure that they are statistically uncorrelated~\cite{sakong2016structure}. We show that RNN-based algorithms can generate trajectory paths as accurately as AIMD simulations, but in a much shorter time. The presented approach treats interactions between atoms as temporary correlations; thus, the problem of predicting the trajectory of atoms reduces to a multi-variate time-series forecasting problem. 

Time-series forecasting has become a groundbreaking research topic over the past couple of years; a significant example is the seq2seq model proposed by Google to make multi-step sequence predictions in order to solve the machine translation problem~\cite{sutskever2014sequence}. An important class of RNN architectures designed for sequence-to-sequence problems is called the encoder-decoder LSTM, which comprises an encoder that receives input signals and returns a fixed-length internal representation as a vector, and a decoder that interprets this vector and uses it to predict the output signal~\cite{wu2016google}. This approach is powerful as not only we can have an input of univariate time-series data, but also, we can encode the multi-variate features of a time-series data-set. Various other types of RNNs have been employed for such predictions such as long short-term memory (LSTM)~\cite{schultz2018prediction}, bidirectional long short-term memory (BLSTM) and mixture density network (MDN) approaches~\cite{zhang2019research, zhao2018applying}, and gated recurrent units (GRU)~\cite{wolski2017probabilistic}. 

Zhao et al.~\cite{zhao2018applying} proposed a BLSTM-MDN approach for three-point shot prediction in basketball games. They have utilized a Python library called Hyperopt~\cite{bergstra2015hyperopt} for hyperparameter self-tuning during model training. Faster convergence rate and more accurate prediction are the superior features of their approach in comparison to CNNs and MDNs. In terms of predicting the trajectories, their proposed model can produce new trajectory samples beyond predicting those stemming from the real data. Another example reported is predicting amazon spot price using a three-layer LSTM-based network, composed of two LSTM layers and one dense neural network layer, by Baughman et al.~\cite{baughman2018predicting}. Their model has revealed a reduction in mean square error (MSE) of 60\% up to 95\% for different training-sets relative to the well-known Autoregressive Integrated Moving Average (ARIMA) model~\cite{1216141}. 

Sagheer and Kotb~\cite{sagheer2019time} employed deep LSTM (DLSTM) recurrent networks for univariate time series forecasting of petroleum production. To find the optimum architecture of DLSTM and optimal selection of hyperparameters, a genetic algorithm was used in their work. For evaluation purposes, the ARIMA model~\cite{1216141}, the Vanilla RNN model~\cite{gulcehre2014learned}, the deep GRU model~\cite{chung2014empirical}, the Nonlinear Extension for linear Arps decline model~\cite{ma2018predicting}, and the Higher-Order Neural Network model~\cite{chakra2013innovative} were employed. Using root mean square error and root mean square percentage error~\cite{hyndman2006another} measures, the proposed deep LSTM approach, outperforms the other standard models.

In the following, we first describe the GRU and LSTM methodologies followed by the computational details of our AIMD simulations for data-set preparation. Next, we propose our two-layer network model for trajectory predictions of atoms in solid-state structures which consists of a GRU or LSTM layer and one dense layer using sigmoid activation function to condense output from the previous layer. We evaluate this network architecture by comparing the accuracy of trajectory predictions against AIMD simulations.
\begin{figure}[ht]
  \centering
\includegraphics[width=\textwidth]{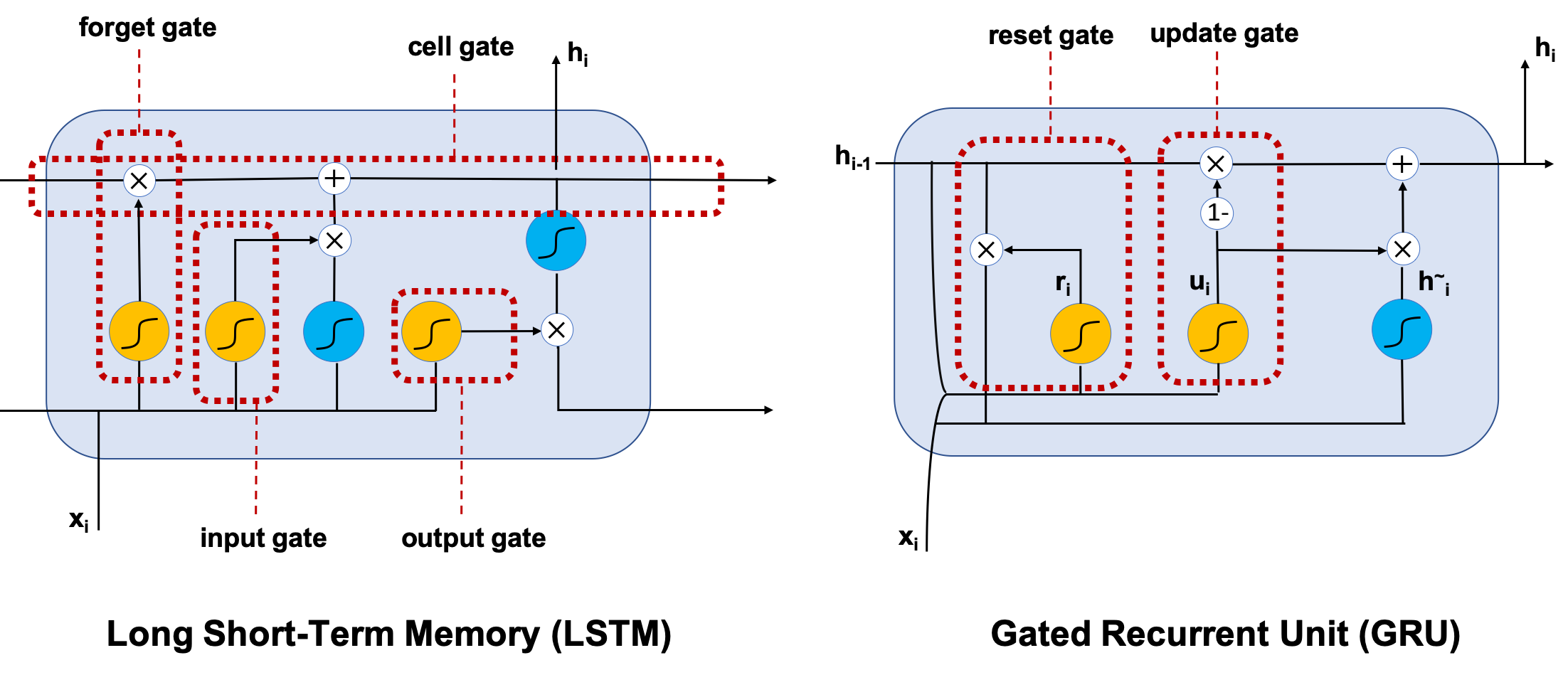}
  \caption{The LSTM and GRU blocks. Yellow and blue circles represent sigmoid and tanh functions, respectively. White circles show point-wise operations.}
  \label{fig:fig1}
\end{figure}
\section{Methodology and Data Preparation}
\label{sec:headings}

\subsection{LSTM/GRU networks}
RNNs receive as input a sequence of signals with no predetermined size limit and utilize their internal states (so-called as memories) for processing the input. The LSTM model, as a special kind of RNN, was proposed by Sepp Hochreiter et al. in 1997~\cite{hochreiter1997long}, to solve the vanishing gradient issue associated with RNNs~\cite{cho2014learning, mandic2001recurrent}. The main problem is, by proceeding to the lower layers thorough back propagation, the gradient or partial derivative of the cost function with respect to the layer’s weight, $ W $, becomes vanishingly small, preventing the weights from updating their values~\cite{hochreiter1998vanishing}. To address this problem, LSTMs employ a gating mechanism, composed of three gates, which give the model permission to pass or forget the data, as shown in Figure~\ref{fig:fig1}~\cite{hochreiter1997long}. The proposed gates establish a new relationship between data and forget previous relationships over a particular time span. Thus, the gating mechanism enables the previous input signals to affect the current signal at a given time while remaining unaffected by those signals far apart~\cite{hochreiter1997long}. Gated Recurrent Unit (GRU) is a form of LSTMs introduced in 2014~\cite{chung2014empirical}. In what follows, we only present the architecture of a GRU which is similar to that of a LSTM model.

In a GRU model, as shown in Figure~\ref{fig:fig1}, a reset gate, r, and an update gate, u, are the main gates. Concatenating the new input with the preceding memory is enabled by the reset gate. The update gate, on the other hand, determines the portion of the preceding memory to be kept. Denote the hidden state of the GRU at time step i by $h_{i}^{'}$; in a GRU structure, the activation $h_{i}^{'}$ at time i is a linear interpolation between the previous activation $h_{i-1}^{'}$ and the candidate activation $(h_{i})$. An update gate $u_{i}$ governs this linear interpolation, and the candidate activation ($h_{i}^{~}$)  depends on a reset gate $r_{i}$. The exact formulation is as follows
\begin{equation}
h_{i}^{'}=(1-u_{i})h_{i-1}^{'}+u_{i}\widetilde{h_{i}},\
\end{equation}
\begin{equation}
u_{i}=\sigma(W_{u}x_{i}+U_{u}h_{i-1}^{'}),\
\end{equation}
\begin{equation}
\widetilde{h_{i}}=tanh(W_{u})x^{i}+U(r_{i}\odot h_{i-1}^{'} ),\
\end{equation}
\begin{equation}
r_{i}=\sigma (W_{r}x_{i}+U_{r}h_{i-1}^{'})\
\end{equation}
Here, vector $x_{i}$ represents the input. The update gate $u_{i}$ in Equation 2 decides how much of each dimension of the hidden state is retained and how much should be updated with the transformed input $x_{i}$ from the current time step.
\begin{figure}[ht]
  \centering
  \includegraphics[width=\textwidth]{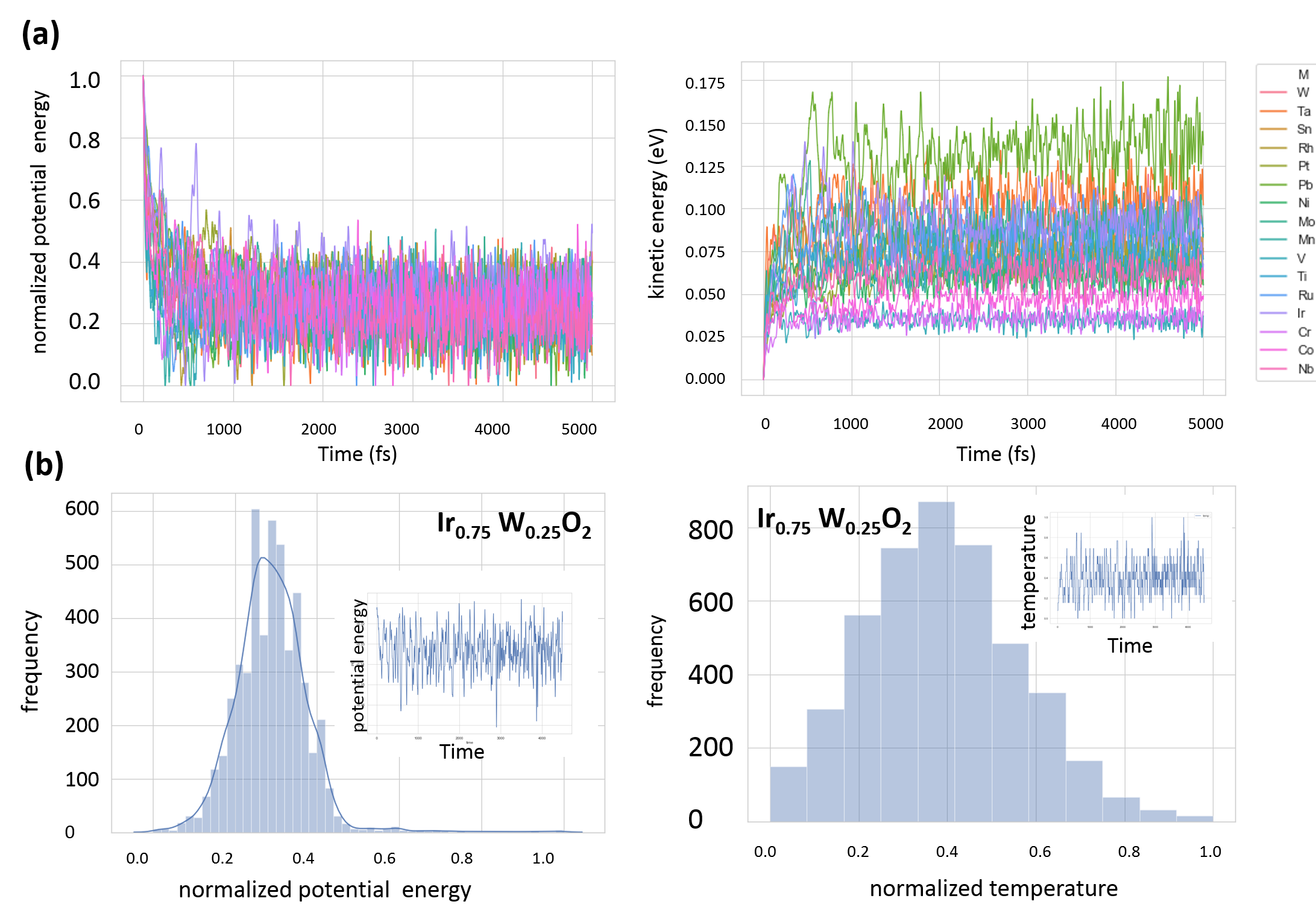}
  \caption{Data prepared based on our AIMD simulations over 5000 fs of the rutile  $ Ir_{0.75}M_{0.25}O_{2} $ (100) surfaces, where M is Ir, Ni, Ti, V, Cr, Co, Nb, Ru, Rh, Sn, Ta, Pt, Pb, Mn, W, Mo. (a) potential energy and kinetic energy profiles for all systems over AIMD simulation (b) distribution of potential energy and temperature for $ Ir_{0.75}W_{0.25}O_{2} $ during the AIMD simulation.}
  \label{fig:fig2}
\end{figure}
\subsection{Data preparation}

Vienna \textit{ab initio} Simulation Package (VASP 5.2) was used to perform the DFT calculations~\cite{kresse1996efficient}. Exchange-correlation effects were incorporated within the generalized gradient approximation (GGA), using the functional by Perdew, Burke, and Ernzerhof (PBE)~\cite{perdew1996generalized}. All calculations were performed based on the Projector Augmented Wave (PAW) method~\cite{blochl1994projector}. According to the PAW method, core electrons were kept frozen and replaced by pseudopotentials (Ir, M, O, H) and valence electrons (Ir: 6s2 5d7; O: 2s2 2p4; H: 1s1) are expanded in a plane wave basis set up to an energy cut-off of 400 eV. Geometry optimization studies were terminated when all forces on ions were less than 0.05 $ eV \AA ^{-1} $.The Brillouin zone was sampled with the Monkhorst Pack  
scheme~\cite{monkhorst1976special} using $ 6\times6\times8 $ k-points. 

For the general raw data, Co, Cr, Ir, Mn, Mo, Nb, Ni, Pb, Pt, Rh, Ru, Sn, Ta, Ti, V and W were substituted for metal M in the $ Ir_{x}M_{1-x}O_{2} $ bulk and the unit-cell was optimized, to account for the strain effects, from which the (100) facet was generated. $ IrO_{2} $ considered in this study has a tetragonal structure with lattice parameters $ a = b = 4.51 $~\AA, $ c = 3.18 $~\AA. AIMD simulations were performed in the microcanonical ensemble using the Verlet algorithm with a time step of 1 fs. The simulations were initialized using configurations obtained from the DFT-based energy minimization. The number of k-points in AIMD simulations was decreased to $ 2\times2\times1 $.  

Trajectory raw data for all atoms were extracted from the VASP output using Visual 
Molecular Dynamics~\cite{humphrey1996vmd} and reshaped for neural network training. No missing data-points were present as expected from AIMD simulations. However, 4 or 5 atoms at the boundaries out of the total 36 atoms in the unit-cell were found to be moving from one side of the cell to the other during the simulation, which was due to the boundary conditions applied in the solid state simulation. We identified these atoms and removed their data from the data-set. 

We used Python 3.5, TensorFlow 1.12.0~\cite{abadi2016tensorflow} and Keras 2.1.6-tf~\cite{chollet2015keras}, pandas 0.24.1~\cite{mckinney2010data}, sklearn~\cite{pedregosa2011scikit}, matplotlib~\cite{hunter2007matplotlib} and seaborn for graphics.

\section{Results and Discussion}

We first generated trajectory data from AIMD simulations over 5000 fs on rutile $ Ir_{0.75}M_{0.25}O_{2} $  (100) surfaces as baseline model system. Figure~\ref{fig:fig2} (a) shows the normalized potential energy and the kinetic energy of 16 different simulated systems for various metals. After about 500 fs, the slab system was equilibrated. Figure~\ref{fig:fig2} (b) shows the distribution of potential energy and temperature for $ Ir_{0.75}W_{0.25}O_{2} $ (100) over the last 4500 fs of simulation time, which follows a Gaussian type distribution. 

Our model consists of a GRU or LSTM layer and a subsequent dense layer to combine output from the previous layer. As input, the RNN-model receives trajectory data for all atoms in the unit-cell. Moreover, we include data for the potential energy profile as well as the temperature fluctuations over the simulation run. As output, in addition to the x, y, z coordinates of the considered atom, we also aim to predict the fluctuations in the potential energy profile. The predictions are then compared with the AIMD calculated values.

\begin{figure}[ht]
  \centering
  \includegraphics[width=\textwidth]{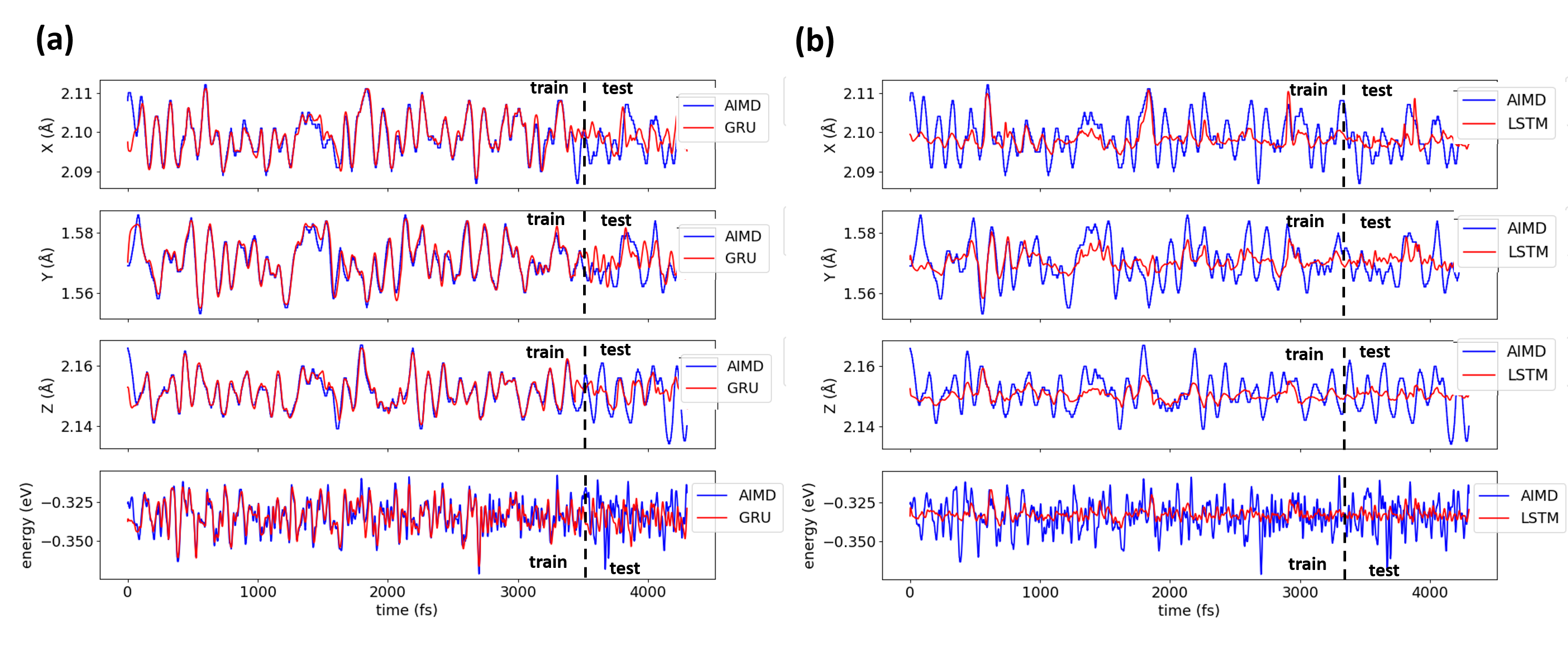}
  \caption{Predictions of the trajectory of the Ir atom in $ Ir_{0.75}W_{0.25}O_{2} $  (100) slab and total potential energy by (a) GRU and (b) LSTM compared to AIMD truth.}
  \label{fig:fig3}
\end{figure}

One test atom for prediction was chosen from the data-set. We resampled the target data for this atom by shifting it negatively by 200 fs (around 5\% of the training-set), meaning that we shifted the target data points to a configuration 200 fs in the past to let the model predict the future trajectory of the test atom. We used 80\% of the AIMD time steps for the training-set and the rest for the test-set. The range of feature values was scaled in the data preprocessing step. To train the RNN, we created batches of shorter sequences of input data selected randomly from the training-set. To prevent overfitting, L2 regularization was performed. RNN was created in Keras~\cite{chollet2015keras} and GRU or LSTM were added to the first layer. This layer returns a sequence of 250 values as input for the dense layer which in turn outputs 4 vectors. The sigmoid activation function was used to output the values between 0 and 1 which is justified as we expect the output values to be in the same range as the training-set. RMSprop optimizer was used to compile the model. The values for the hyperparameters used are shown in Table~\ref{tab:table}. 
\begin{table}
 \caption{List of hyperparameters}
  \centering
  \begin{tabular}{llll}
    \toprule
    \cmidrule(r){1-2}
    Hyperparameters     & Min     & Max & Default Values) \\
    \midrule
    Learning rate & $10^{-4}$  & $5\times10^{-3}$  & $10^{-3}$    \\
    Batch size     & 128 & 512  & 256   \\
    Number of units for GRU and LSTM     & 200       & 300 & 250  \\
        Sequence length & 50  & 200  & 150    \\
    Epochs     & 20 & 20  & 20    \\
    Steps per epoch     & 100       & 200 & 100  \\
            L2 regularization parameter & 0.0001  & 0.001 & 0.001    \\
    Train-test splitr     & 80\%-20\% & 90\%-10\%    & 80\%-20\%     \\
    \bottomrule
  \end{tabular}
  \label{tab:table}
\end{table}
We used mean squared error (MSE) as loss function which required to be minimized during the model training:
\begin{equation}
MSE = \frac{\sum_{i=1}^{n} (y_{i}-y_{i}^{p})^{2}}{n}\
\end{equation}
Figure~\ref{fig:fig3} shows the predicted output sequences, namely x, y, z coordinates and the energy profile of our GRU- and LSTM-based models compared with the ground truth from AIMD simulation. As shown for the training-set, both models predicted the fluctuations of the coordinates with high accuracy. This was expected as the train-data was seen by the models during the training phase. The prediction, however, is not accurate for the first few time steps because the models have not received enough input. Likewise, for the potential energy the fluctuations were also predicted well on the training data. In addition, both models’ predictions were reasonably accurate on the test-set, although some peaks were not matched with the AIMD trajectories and there seems to be a small deviation. 

\begin{figure}[ht]
  \centering
  \includegraphics[width=\textwidth]{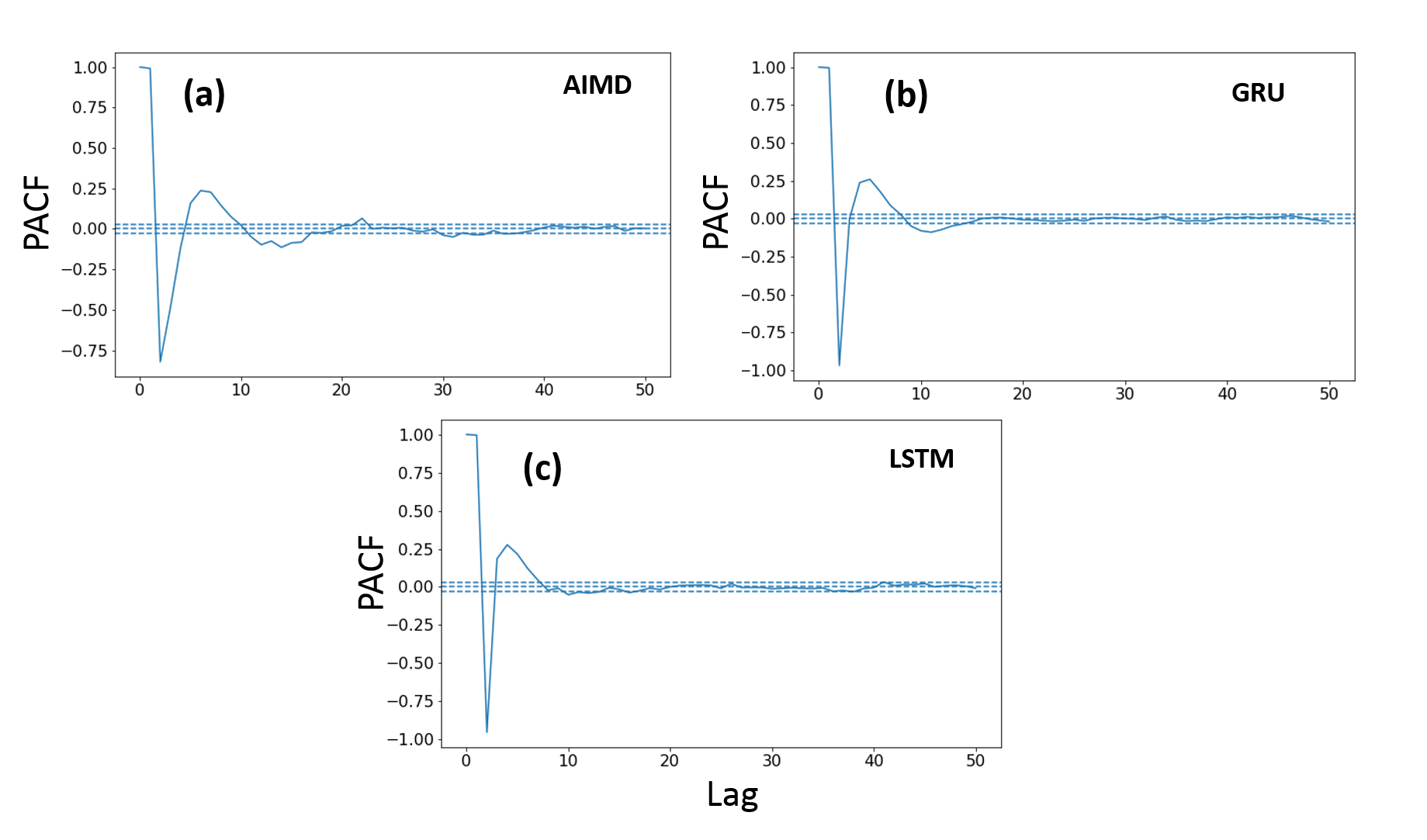}
  \caption{Partial autocorrelation function for the potential energy profile obtained from AIMD, GRU and LSTM.}
  \label{fig:fig4}
\end{figure}

Figure~\ref{fig:fig4} compares the generated partial autocorrelation function (PACF) of the potential energy calculated with AIMD simulations with those predicted by GRU and LSTM models. In time-series, the partial autocorrelation indicates the correlation between an instance with those at prior time steps having the relationships of intervening instances removed. Mathematically, partial autocorrelation function values are the coefficients of an autoregression on lagged values of the time-series as given by,
\begin{equation}
Y_{t}=a_{0}+a_{1} Y_{t-1}+...+a_{N} Y_{t-N}\
\end{equation}
As shown in Figure~\ref{fig:fig4}, the PACF from AIMD simulation and the models indicate that after around 20 fs the energy values become uncorrelated. This particularly shows the ability of GRU and LSTM models to predict uncorrelated configurations which can be used directly for the calculations of thermodynamic properties.  

\begin{figure}[ht]
  \centering
  \includegraphics[width=\textwidth]{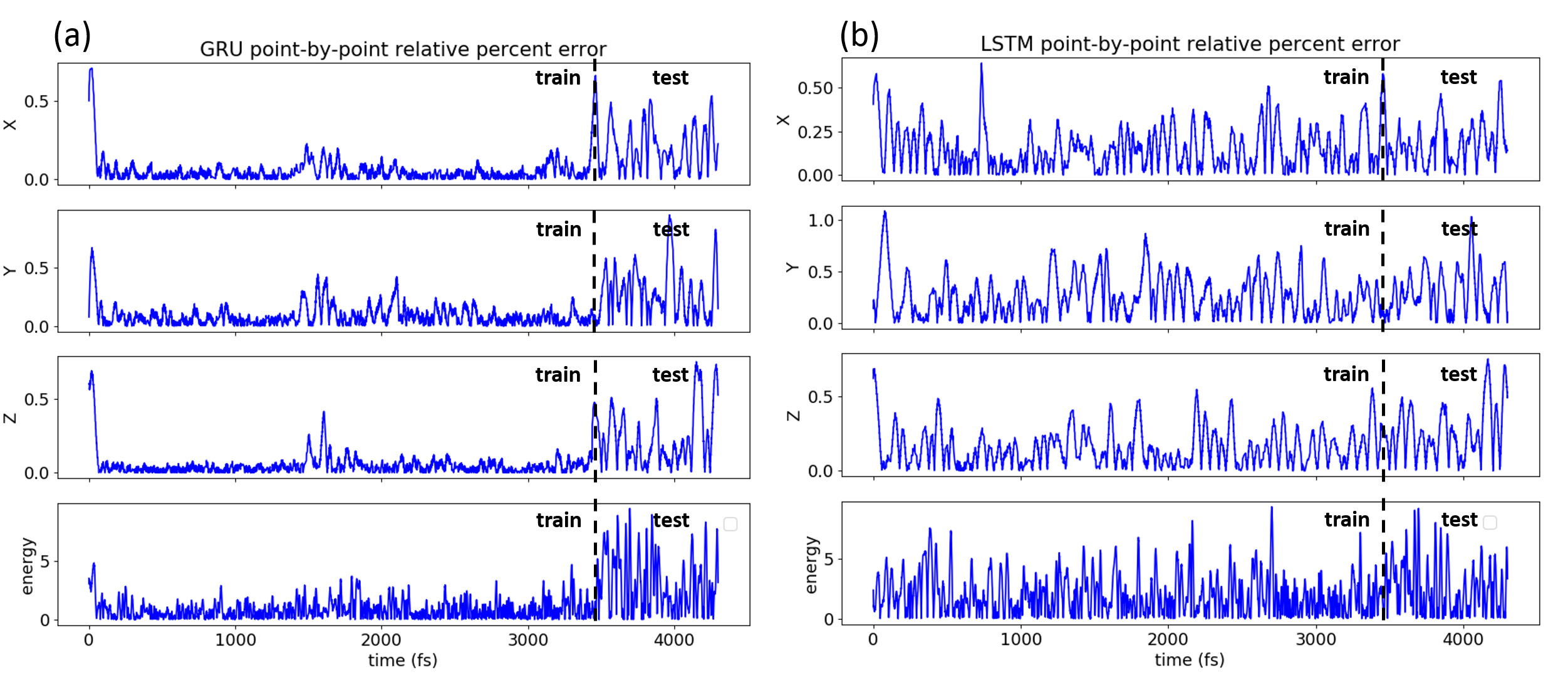}
  \caption{Absolute percentage error of the two models}
  \label{fig:fig5}
\end{figure}

In Figure~\ref{fig:fig5}, we calculated the point-by-point absolute percentage error $((|y_
{i}-y_{i}^{p} |)/y_{i} \times100)$ of GRU and LSTM models against the AIMD simulation. For both the training and test sets, the percentage error of the GRU model for x, y, z coordinates were found to be less than 0.9\%, with a mean error of around 0.1\%.  The mean percentage error for potential energy predicted by GRU was 1.3\%, standard deviation was 1.5\% and the maximum error was found as 9.4\%. Regarding the LSTM model, the maximum percentage error for the coordinates was found as around 1\% with the mean error and standard deviation values of around 0.1-0.2\%. The calculated mean error for the potential energy predicted by LSTM was 2.1\% with standard deviation of 1.7\% and maximum value of 9.3\%. Even though the calculated error shows slightly more accurate predictions for the GRU model, LSTM seems to be more robust in terms of dealing with the overfitting issue as the error values for both train and test sets lie in pretty much the same range.

\begin{figure}[ht]
  \centering
  \includegraphics[width=\textwidth]{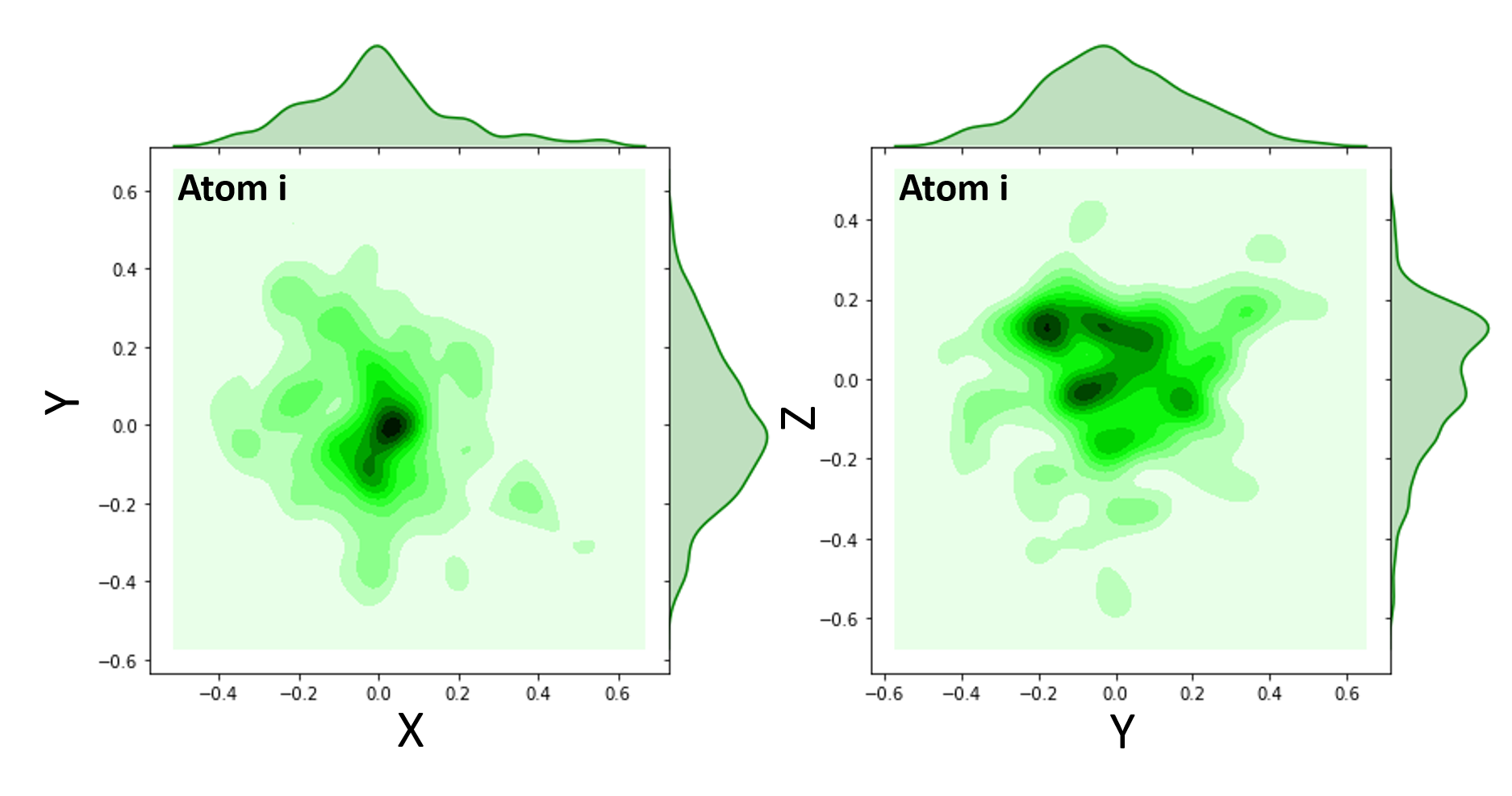}
  \caption{Density plot of coordinates of (a) all atoms (b) randomly chosen atom i, over the AIMD simulation run.}
  \label{fig:fig6}
\end{figure}

Figure~\ref{fig:fig6} shows the spatial distribution of the atoms in the unit-cell over the sampling run, which approximately follows a normal shape. Moreover, the temperature fluctuation and potential energy profile used as input signals were also found to be of Gaussian type. Therefore, it is not surprising to observe highly accurate predictions by the GRU and LSTM models on this data-set, as such algorithms tend to learn more easily from this kind of distribution. 

Overall this study suggests that statistical sampling of a solid-state system can be swiftly performed with GRU/LSTM, thus, RNN-based models can be potentially employed as complementary tools AIMD simulations in order to markedly reduce computational costs. For future studies, these models will be tested on more complicated systems, simulating how a system behaves out of equilibrium, e.g. during relaxation times, phase transitions or defect formation. 

\section{Conclusions}
In this work, we explored two types of recurrent neural networks, namely gated recurrent unit and long short-term memory networks, for the fast prediction of trajectory and potential energy profiles in \textit{ab initio} molecular dynamics simulations. We found that these algorithms can learn from the distribution of the coordinates of atoms over the simulation run. We treated the problem as a multi-variate time-series and referred the interactions between the atoms to temporary correlations. Our model is compared against the baseline AIMD simulation. Findings suggest that both GRU and LSTM networks can be considered as alternative models to AIMD.

\section{Acknowledgments}
This research was enabled in part by support from WestGrid (www.westgrid.ca) and Compute Canada Calcul Canada (www.computecanada.ca) which provided 4 x NVIDIA P100 Pascal (16G HBM2 memory) GPU for this research.

\bibliographystyle{unsrt}  
\bibliography{references}  

\end{document}